%% file: Vass_valluri_etal.tex
\documentclass[useAMS,usenatbib]{mn2e}
\input{mycommands}

\usepackage{times} 
\usepackage{graphicx}
\usepackage{rotating}
\usepackage{epsfig}
\usepackage{amsmath}

\def\funits{M_\odot{\rm kpc}^{-3}({\rm km s}^{-1})^{-3}\,}
\def\fiestas{FiEstAS\,}
\def\enbid{EnBiD\,}

\title[Evolution of phase-space density in LCDM halos]{Evolution of the Dark
Matter Phase-Space Density Distributions of $\Lambda$CDM Halos}

\author[Vass~et~al.]{Ileana M. Vass$^{1,2}$, Monica
Valluri$^{2,3}\thanks{direct queries to: mvalluri@umich.edu (MV)}$, Andrey V.
Kravtsov$^{2,4,5}$, Stelios Kazantzidis$^{6}$ \\
$^{1}$ Department of Astronomy University of Florida, Gainesville, FL 32611,
USA \\
$^{2}$ Kavli Institute for Cosmological Physics, The University of Chicago,
Chicago, IL 60637, USA,  \\
$^{3}$ Department of Astronomy, University of Michigan,  Ann Arbor, MI 48109,
USA, ({\tt mvalluri@umich.edu})\\
$^{4}$ Enrico Fermi Institute, The University of Chicago, Chicago, IL 60637,
USA \\
$^{5}$ Dept. of Astronomy \& Astrophysics, The University of Chicago, 5640 S.
Ellis Ave., Chicago, IL 60605\\
$^{6}$ Center for Cosmology and Astro-Particle Physics, The Ohio State
University, Columbus, OH 43210, USA}
\begin{document}

\date{Accepted ------. Received ------; in original form ---}

\pagerange{\pageref{firstpage}--\pageref{lastpage}} \pubyear{2008}

\maketitle

\label{firstpage}
\begin{abstract}  

We study the evolution of phase-space density during the hierarchical structure
formation of $\Lambda$CDM halos.  We compute both a spherically-averaged
surrogate for phase-space density ($Q = \rho/\sigma^3$) and the coarse-grained
distribution function $f(\mathbf{x},\mathbf{v})$ for dark matter particles that
lie within $\sim 2$ virial radii of four Milky-Way-sized dark matter halos. The estimated $f(\mathbf{x},
\mathbf{v})$ spans over four decades at any radius. Dark matter particles that end up within two virial
radii of a Milky-Way-sized DM halo at $z=0$ have an approximately Gaussian
distribution in $\log(f)$ at early redshifts, but the distribution becomes
increasingly skewed at lower redshifts. The value $f_{\rm peak}$ corresponding
to the peak of the Gaussian decreases as the evolution progresses and is well
described by $f_{\rm peak}(z) \propto (1+z)^{4.3 \pm 1.1}$.  The decrease is
due both to the dynamical mixing as the matter accreted by halos is virialized,
and due to the overall decrease of space-density of the unprocessed material
due to expansion of the Universe.  The highest values of $f$, (responsible for the skewness
of the profile) are found at the centers of dark matter halos and subhalos, where $f$ can be an order of
magnitude higher than in the center of the main halo.  We confirm that $Q(r)$ can be described by a power-law with the slope of $\beta
= -1.8 \pm 0.1$ over 2.5 orders of magnitude in radius and over a wide range of
redshifts. This $Q(r)$ profile likely reflects the distribution of entropy ($K
\equiv \sigma^2/\rho_{\rm dm}^{2/3} \propto r^{1.2}$), which dark matter
acquires as it is accreted onto a growing halo. The estimated $f(\mathbf{x},
\mathbf{v})$, on the other hand, exhibits a more complicated behavior. Although
the median coarse-grained phase-space density profile $F(r)$ can be
approximated by a power-law, $\propto r^{-1.6 \pm 0.15}$, in the inner regions
of halos ($<0.6 \; r_{\rm vir}$), at larger radii the profile flattens
significantly. This is because phase-space density averaged on small scales is
sensitive to the high-$f$ material associated with surviving subhalos, as well
as relatively unmixed streams resulting from disrupted subhalos, which
constitute a sizeable fraction of matter at large radii.

\end{abstract}


\bigskip
}
\section{Introduction}
\label{sec:introduction}

Cosmological $N$-body simulations of the formation of structure in the
Universe allow us to reconstruct how gravitationally bound objects
like galaxies and clusters form and evolve. These simulations have
shown that, despite the seemingly complex hierarchical formation
history of dark matter (DM) halos, the matter density distributions of
cosmological DM halos are described by simple 3-parameter profiles
\citep{NFW, Metal2006}.

The study of the evolution of the phase-space distribution function (DF) of a
collisionless dynamical system (composed of stars or DM or both) is fundamental
to understanding its dynamical evolution, since a collisionless system is
completely described by its phase-space density distribution (otherwise called
the fine-grained DF $\mathbf{f}(\mathbf{x}, \mathbf{v})$). For an isolated
collisionless system, $\mathbf{f}(\mathbf{x}, \mathbf{v}, t)$ is fully
described by the Collisionless Boltzmann Equation (CBE) (sometimes called the
Vlasov equation), which is a continuity equation in phase-space. A consequence
of the CBE is that the fine-grained DF $\mathbf{f}$ is always conserved.  In
addition, $\mathbf{V}(\mathbf{f})d\mathbf{f}$, the volume of phase-space
occupied by phase-space elements whose density lies between
$(\mathbf{f},\mathbf{f}+d\mathbf{f})$, is also conserved.  However, the
conservation of the fine-grained $\mathbf{f}(\mathbf{x}, \mathbf{v})$ and
$\mathbf{V}(\mathbf{f})d\mathbf{f}$ is not a very useful property for
understanding the evolution of real systems, since in practice it is only
possible to compute an average of $\mathbf{f}$ over a finite volume of
phase-space. This average is referred to as the coarse-grained DF
$f(\mathbf{x}, \mathbf{v})$, and the associated volume density is $V(f)$.  The
evolution of the coarse-grained DF is governed by {\em Mixing Theorems}
\citep{BT, tremaine_henon_lyndenbell_86, mathur_88}, which state that if the
coarse-grained DF is bounded, then processes that operate during the relaxation
of collisionless systems (e.g.  phase mixing, chaotic mixing, and the mixing of
energy and angular momentum that accompanies violent relaxation) result in a
decrease in all convex functions of the coarse-grained phase-space density.  In
particular, the coarse-grained phase-space density in a region of phase-space
is expected to decrease with time, because at any point in phase-space, both
low and high phase-space density regions get highly mixed \citep{BT,
tremaine_henon_lyndenbell_86}. In other words, mixing reduces $f$ such that, at
any time, $ f \; < \; f_{\rm max}$, the maximum phase-space density at the
start. In addition, there are simple relationships that exist between the
initial $\mathbf{V}(\mathbf{f})$ and the coarse-grained $V(f)$ at a later time:
$V(f)$ is always greater than $\mathbf{V}(\mathbf{f})$ \citep{mathur_88}.

Recently, \citet{dehnen2005} stated and proved a new Mixing Theorem
that states that the excess-mass function $D(f)$, which is the mass of
material with coarse-grained phase-space density greater than a value
$f$, always decreases due to mixing, for all values of $f$.  The
excess-mass function is additive so that the excess mass of a
combination of non-overlapping systems is the sum of their individual
excess masses. He showed using this theorem that steeper cusps are
less mixed than shallower ones, independent of the details of the DF
or density profile, and this implies that a merger remnant cannot have
a cusp that is either steeper or shallower than the steepest of its
progenitors. This theorem has powerful implications for the evolution
of halos with cuspy power-law central phase-space density profiles via
mergers, once cuspy profiles have been formed.

\citet{taylorN01} have introduced a surrogate quantity, $Q(r) =
\rho(r)/\sigma(r)^{3}$, where $\rho(r)$ is the configuration-space density
averaged in spherical shells, and $\sigma(r)$ is the velocity dispersion of
dark matter particles averaged in a spherical shell centered at the radius $r$.
Note that, although $Q(r)$ has dimensions of phase-space density, it is
definitely {\it not} true coarse-grained phase-space density, because it is
constructed out of separately computed configuration-space density and velocity
dispersion in arbitrarily chosen spherical shells. Consequently, it is not
expected to obey any {\em Mixing Theorems} \citep{dehnen_mclaughlin_05} and is
generally difficult to interpret in terms of phase-space density. Nevertheless,
it is an intriguing quantity, because \citet{taylorN01} show that $Q(r)$ is well
approximated by a single power-law, $Q \propto r^{-\beta} $ with $\beta \approx
1.874$ over more than 2.5 orders of magnitude in radius for CDM halos
universally, regardless of mass and background cosmology \citep[see
also][]{rasia_etal04,ascasibar_etal04}. Recent work, however,  indicates that
$Q(r)$ profiles are somewhat sensitive to the amount of substructure and the
slope of the power spectrum \citep{wang_white08,knollman_etal08}.

While power-law $Q(r)$ profiles are by no means universal for
self-gravitating systems in dynamical equilibrium \citep{barnes06},
power-law profiles with exactly the same slope were first shown to
arise in simple self-similar spherical gravitational infall models
\citep{bertschinger_85} pointing to a possible universality in the
mechanisms that produces them. \citet{austin05} extended the early
work of \citet{bertschinger_85} using semi-analytical extended
secondary infall models to follow the evolution of collisionless
spherical shells of matter that are initially set to be out of
dynamical equilibrium and are allowed to move only radially.  They
concluded that the power-law behavior of the final phase-space density
profile is a robust feature of virialized halos, which have reached
equilibrium via violent relaxation. They used a constrained Jeans
equation analysis to show that this equation has different types of
solutions and that it admits a unique periodic solution which gives a
power-law density profile with slope $\beta = 1.9444$ (comparable to
the numerical value of $\beta = 1.87$ obtained by \citet{taylorN01}).

A recent study of \citet{hoffman07} followed the evolution of $Q(r)$
profiles in cosmological DM halos in constrained simulations designed
to control the merging history of a given halo. These authors showed
that, during relatively quiescent phases of halo evolution, the density
profile closely follows that of a NFW halo, and $Q(r)$ is always well
represented by a power-law $r^{-\beta}$ with $\beta = 1.9\pm
0.1$. They showed that $Q(r)$ deviates from power-law most strongly
during major mergers but recovers the power-law form thereafter
\citep[this is consistent with our findings using a series of
controlled merger simulations, ][]{vass_etal_08a}.  More recently,
\citep{wojtak_etal08} demonstrated (using a varient of the \fiestas
code) that the DF of $\Lambda$CDM halos of mass $10^{14}-
10^{15}~M_{\odot}$ can be separated into energy and angular momentum
components and proposed a phenomenological model for spherical
potentials. They showed that their model DF was a good match to the
N-body DFs and reproduced the power-law behavior of $Q(r)$.  Another
study of relevance is the work of \citet{pei_fr_pac07}, which defined
a global value for the phase-space density, $Q$, which they found
decreases rapidly with time. Despite the insights obtained in these studies,
the origin for such universality of $Q(r)$ is not yet understood, and
the question of how this quantity relates to the true coarse-grained 
phase-space density has not yet been thoroughly investigated, which
motivates our study in which we focus on the evolution of {\it both}
$Q(r)$ and phase-space density evolution using a set of self-consistent
cosmological simulations of halo formation.  

In the last four years, three independent numerical codes to compute
the true coarse-grained phase-space density from $N$-body simulations
have been developed \citep{Arad04, ascasibar_binney05,
shar_stein06}. The first study of coarse-grained $f$ showed that it
has its highest values at the centers of DM halos and subhalos. They
also showed that the volume density of phase-space $V(f)$ within
individual cosmological dark matter halos at $z=0$ has a power-law
profile over nearly four orders of magnitude in $f$ \citep{Arad04}.
The main criticism of this code (which is based on the Delunay
tesselation of phase-space) is that it is not metric free. In this
paper, we compare results obtained with two other codes ``\fiestas''
\citep{ascasibar_binney05} and ``\enbid'' \citep{shar_stein06}, which
are both fast and metric free\footnote{ An even more sophisticated
numerical method has recently been presented by
\citet{vogelsberger_etal08} for calculating the fine-grained phase-space
structure of dark matter distributions derived from cosmological
simulations. This code has the potential to identify fine-scale
structure such as caustics in phase-space and the phase-space
structure of tidal streams in the Milky-Way halo. Its ability to
estimate the fine-grained $\mathbf{f}(\mathbf{x}, \mathbf{v})$ makes
it useful for understanding non-equilibrium systems and for making
accurate predictions for direct DM searches.}, to study the
evolution of the course-grained phase-space density.

In an accompanying paper \citep{vass_etal_08a}, we present analysis of
phase-space and $Q(r)$ evolution during major mergers between CDM-like
spherical halos.  We show that major mergers (those which are the most
violent and therefore likely to result in the greatest amount of
mixing in phase-space) preserve spherically-
averaged phase-space density profiles (both $Q(r)$ and $F(r)$) out to
the virial radius $r_{\rm vir}$ of the final remnant. Mergers between
identical but non-NFW halos, i.e. halos with central cusps
significantly shallower than NFW ($\gamma = 0.2$) or steeper than NFW
($\gamma = 1.7$), also retain their original phase-space density
profiles.  Mergers between a halo with a shallow cusp and a steep cusp
produce halos with inner phase-space density profiles that are
indistinguishable from that of the steepest cusp in the pair of
progenitors.  These results were anticipated by the work of
\citet{dehnen2005} and are a consequence of the Mixing Theorem for
cusps, which requires that the excess mass function of collisionless
systems always decreases. Thus we confirm the prediction
(\citet{dehnen2005}) that the phase-space density profiles of DM halos
are extremely robust, and in particular, the steepest central cusp
always survives.

To understand the robustness of profiles outside the center cusp
\citet{vass_etal_08a}, we drew on recent work by \citet{valluri07}
that shows that matter in equally-spaced radial shells is
redistributed during the merger is such a way that only about 20\% of
the matter in the central cusp is ejected during the merger to radii
extending out to about two scale radii. For all other radii, roughly 15\%
by mass in each radial shell is redistributed uniformly with radius.
Further, nearly 40\% of the mass of the progenitor halos
lies beyond the formal virial
radius of the remnant \citep{kazantzidis_etal06, valluri07}, and this matter originates roughly uniformly
from each radial interval starting from about three scale radii from the
center.  While major mergers, such as those studied by
\citet{dehnen2005,vass_etal_08a}, are instructive, since they represent
the most extreme form of mixing, they are not the major mode of mass
accretion in the Universe. In addition, since experiments with major
mergers demonstrate that pre-existing power-law profiles are
preserved, they shed little light on the formation of the
profiles. Our primary goal is to gain a better understanding of the
origin of the power-law phase-space density profiles seen in
cosmological DM simulations.

In this paper we investigate the evolution of the coarse-grained phase-
space density in the formation and evolution of four Milky-Way-sized
halos in a $\Lambda$CDM cosmology with cosmological parameters:
$(\Omega_{m},\Omega_{\Lambda},h,\sigma_{8})=(0.3,0.7,0.7,0.9)$.  In
\S~\ref{sec:numerical} we summarize the cosmological N-body
simulations analyzed in this study as well as the numerical methods
used to obtain coarse-grained phase-space densities. (A detailed
comparison of the two codes used (\fiestas and \enbid) is presented in
the Appendix, where we give reasons for our choice of \enbid, with
a $n=10$ smoothing kernel).  In \S~\ref{sec:onehalo} we describe the
properties of the spherically-averaged quantity
$Q(r)$ as well as the the coarse-grained DF for one Milky-Way-sized DM
halo at $z=0$, as well as the evolution of the phase-space DF of this
halo from $z=9$ to the present. In \S~\ref{sec:otherhalos} we compare
these results for three additional Milky-Way-sized halos. In
\S~\ref{sec:origin} we present an interpretation of the power-law profiles seen in $Q(r)$ and discuss the implications of observed coarse grained distribution function  in a
cosmological context. \S~\ref{sec:conclude} summarizes the results of
this paper and concludes.

\bigskip
\section{Numerical  Methods}
\label{sec:numerical}

The simulations analyzed in this paper are described in greater detail
in the works of \citet{tumultuos} and \citet{fossils}. The simulations
were carried out using the Adaptive Refinement Tree $N$-body code
(ART, \citet{art}). The simulation starts with a uniform $256^{3}$
grid covering the entire computational box.  This grid defines the
lowest (zeroth) level of resolution. Higher force resolution is
achieved in the regions corresponding to collapsing structures by
recursive refining of all such regions by using an adaptive refinement
algorithm. Each cell can be refined or de-refined individually. The
cells are refined if the particle mass contained within them exceeds a
certain specified value. The grid is thus refined to follow the
collapsing objects in a quasi-Lagrangian fashion.

Three of the galactic DM halos were simulated in a comoving box of 25
$h^{-1}$ Mpc (hereafter L25); they were selected to reside in a
well-defined filament at $z = 0$. Two halos are neighbors, located 425
$h^{-1}$ kpc from each other.  The third halo is isolated and is
located 2 Mpc away from the pair.  Hereafter, we refer to the isolated
halo as G1 and the halos in the pair as G2 and G3.  The virial masses
and virial radii for the halos studied are given in Table 1 in
\citet{tumultuos}. The virial radius (and the corresponding virial
mass) was chosen as the radius encompassing a mean density of $\sim 200$
times the mean density of the Universe.  The masses of the DM halos
are well within the range of possibilities allowed by models for the
halo of the Milky Way galaxy \citep{klypin02}.  The simulations
followed a Lagrangian region corresponding to the sphere of radius
equal to two virial radii around each halo at $z=0$.  This region was
re-sampled with the highest resolution particles of mass $m_{\rm
p}=1.2\times 10^6h^{-1}{\rm M_{\odot}}$, corresponding to $1024^3$
particles in the box, at the initial redshift of the simulation
($z_{\rm i}=50$).  The maximum of ten refinement levels was reached in
the simulations corresponding to the peak formal spatial resolution of
$150$ comoving parsec.  Each host halo is resolved with $\sim 10^6$
particles within its virial radius at $z=0$.

The fourth halo was simulated in a comoving box of 20 $h^{-1}$ Mpc box
(hereafter L20) and it was used to follow the Lagrangian region of
approximately five virial radii around the Milky Way-sized halo with
high resolution.  In the high-resolution region the mass of the dark
matter particles is $m_{\rm p}=6.1\times 10^5h^{-1}{\,\rm}M_{\odot}$,
corresponding to effective $1024^3$ particles in the box, at the
initial redshift of the simulation ($z_{\rm i}=70$).  As in the other
simulation, this run starts with a uniform $256^3$ grid covering the
entire computational box. Higher force resolution is achieved in the
regions corresponding to collapsing structures by recursive refining
of all such regions using an adaptive refinement algorithm. Only
regions containing highest resolution particles were adaptively
refined.  The maximum of nine refinement levels was reached in the
simulation corresponding to the peak formal spatial resolution of
$150h^{-1}$ comoving parsec.  The Milky Way-sized host halo has the
virial mass of $1.4\times 10^{12}h^{-1}\,\rm M_{\odot}$ (or $2.3$
million particles within the virial radius) and virial radius of
$230h^{-1}{\rm kpc}$ at $z=0$.

The peak spatial resolution of the simulations determines the minimum
radius to which we can trust the density and velocity profiles. In the
following analysis, we only consider profiles at radii at least eight
times larger than the peak resolution of the simulations (i.e.,
$\approx 0.5-0.6h^{-1}$~comoving kpc or $r\approx 0.004\, r_{\rm
vir{(z=0)}}$).

\bigskip
\subsection{Phase-space density estimators}
\label{sec:estimators}

In this paper, we will focus on the time evolution of the spherically averaged 
phase-space density  $Q(r)$ as well as the coarse-grained phase-space
DF $f(\mathbf{x}, \mathbf{v})$ during the formation of  the
Milky-Way sized DM halos in $\Lambda$CDM cosmological simulations described
above.  

In the last four years three independent numerical codes to compute
the coarse-grained phase-space density from $N$-body simulations have
been developed \citep{Arad04, ascasibar_binney05, shar_stein06}. These
techniques differ in the scheme used to tesselate 6-dimensional phase
space as well as in the density estimators they use.  In our analysis,
we use the FiEstAS algorithm of \citet{ascasibar_binney05} and the
EnBid algorithm of \citet{shar_stein06} due to their speed and the
metric free nature.  The {\fiestas} algorithm is based on a repeated
division of each dimension of phase-space into two regions that
contain roughly equal numbers of particles.  \enbid
\citep{shar_stein06} closely follows the method used by \fiestas, but
improves upon it to allow for more accurate computation of $f$ at high
phase-space densities.  In \enbid the space is tesselated along the
dimension having minimum entropy (and maximum information) each step
of the computation.  This scheme optimizes the number of divisions to
be made in a particular dimension and extracts maximum information
from the data by using a minimum entropy criterion based on the {\it
Shannon entropy} that achieves much greater accuracy at measuring the
phase-space density when it is high.  \citet{shar_stein06} showed that
\enbid with a kernel which includes 10 nearest neighbors ($n=10$)
about each point was able to recover analytic phase-space density
profiles to nearly 3-4 decades higher values of $f$ than \fiestas.

For our halos at $z=0$, the results we obtained with the different
codes were completely consistent with those obtained by
\citet{shar_stein06}. The deviations between the estimates obtained
with the different codes and various parameters were significantly
higher at higher redshift - where comparisons with analytic estimates
are not available. We refer the reader to the Appendix for a detailed
comparison between estimated values of $f$ using the two codes at
various redshifts. In the rest of this paper we present results
obtained with \enbid ($n=10$ kernel), the parameters prefered by
\citet{shar_stein06}.

\begin{figure}
\centerline{\psfig{figure=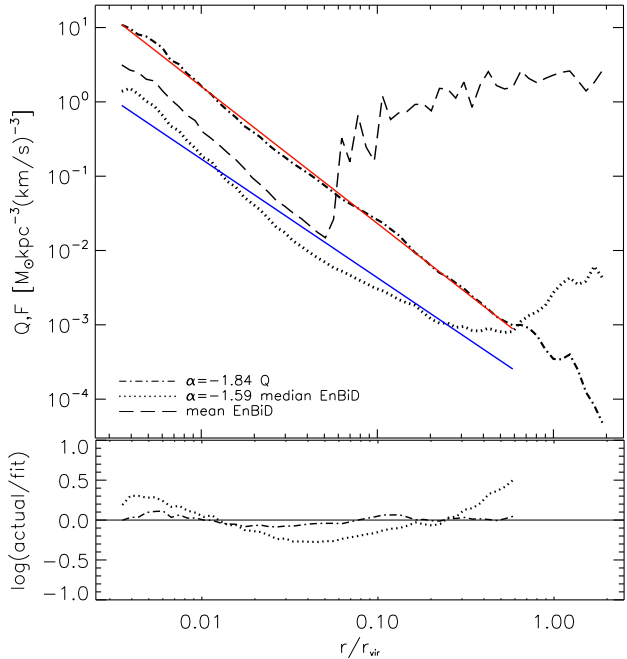,width=8.5cm}}
\caption{$Q(r)$ and $F(r)$ in 100 spherical radial bins about the
most-bound-particle in halo G1 from the L25 simulation at $z$=0.  Top
panel: $Q$ (dot-dashed curve). Power-law fit $Q \propto r^{-1.84 \pm
0.012}$ is given by solid red line.  $F(r)$ from \enbid \, (with $n=10$
smoothing kernel (dotted curve)). Mean values of
$\mathbf{f}$ estimated by \enbid ($n=10$)are given by long-dashed curve is
noiser due to subhalos. A power-law fit to $F(r) \propto r^{-1.59 \pm
0.054}$ is given by solid blue line.  Residuals to the two power-law
fits are shown in the bottom panel.
\label{fig:qandfz0}}
\end{figure}

Figure~\ref{fig:qandfz0} {\it Top} shows the spherically averaged
quantity $Q =\rho/\sigma^{3}$ (dot-dashed curve); the mean of
$f(\mathbf{x}, \mathbf{v}$) in spherical bins (dashed curve) and the
median DF in spherical bins (dotted curve).  As we will show in
Figure~\ref{fig:rf_L25_cont}, the large fluctuations in the mean value
of $f$ (dashed curve) beyond $0.1r_{\rm vir}$ are due to the presence
of substructure, which can have extremely high central values of $f$.
The median value of $f$ (hereafter represented by $F$) is much
smoother, being less sensitive to the large range (nearly eight orders
of magnitude) in $f$ at each radius.  In what follows, we will use the
median $F$, computed in concentric radial bins centered on the most
bound particle in the main halo, since it is less sensitive to
substructure.  $Q(r)$ is well fitted by a power-law: $Q \propto
r^{-1.84\pm 0.012}$ over the radial range $[0.004$--$0.6] \, r/r_{\rm
vir}$ (thin red line).  A power-law is also fitted to $F(r) \propto
r^{-1.59 \pm 0.054}$ (blue line) over the same radial range. The
bottom panel shows the residuals of the fits $F$ ($\log(F/F_{\rm
fit})$) and $Q$ ($\log(Q/Q_{\rm fit})$). This plot shows that while
$Q(r)$ is an extremely good power-law, the median $F(r)$ is only
approximately power-law over the same radial range. A similar result 
was obtained in a much higher resolution simulation by \citet{stadel_etal08}, 
 a study which appeared as we were preparing this paper for publication.

\bigskip
\section{Evolution of phase-space density with redshift}
\label{sec:onehalo}

\citet{hoffman07} studied the phase-space density profiles of a dark
matter halo by tracking $Q(r)$ for material within the formal virial
radius at each redshift.  They found that the virialized material
within this radius has an approximately power-law from with a constant
power-law index of $-1.9\pm 0.05$ at all redshifts from $z=5$ to the present.

In this paper we follow a slightly different approach since our
objective is to understand the evolution of the true coarse-grained
phase-space density distribution. We track the evolution of
phase-space density, by tracing backwards in time, all the material
that lies inside the virial radius at $z=0$. This will allow us to
understand how the initially high phase-space density material that
lies outside virialized systems at high redshift falls in and
undergoes mixing and how the resultant mixing preserves the
phase-space density profiles as a function of redshift.

We identified Milky-Way sized DM halos at $z=0$ in our cosmological
simulations and identify all the particles that lie inside twice the
virial radius at $z=0$.  We then tracked these particles back to
$z=9$. Our simulations do proceed back in time to even higher
redshifts, however we do not analyze them here because the mass
resolution at higher redshifts does not allow us to resolve the
evolution of most of the objects beyond this epoch.  In total there
were 1.6$\times 10^{6}$ particles within 2 virial radii of the G1
halo, which is the halo that is most isolated at $z=0$. Two other
halos (G2 and G3) from the L25 simulation, are closer to each other at
$z=0$ (1.5 virial radii apart), and thus, we only track back particles
within one virial radius of the centers of these two halos.  The
particles in the high resolution simulation L20 were selected and
tracked in the same way as for the G1 halo, and at $z=0$ there were
3$\times 10^{6}$ particles inside twice the virial radius.  The
position and velocity data for all the particles identified as
belonging to a given halo at $z=0$ were analysed in physical
coordinates (distances are always in kpc and velocities in $\rm {km
{s}^{-1}}$).  In this section, we will present the results obtained
for the G1 halo in the L25 simulation. A comparison with results for
the other three halos is made in \S~\ref{sec:otherhalos}.

\begin{figure}
\centerline{\psfig{figure=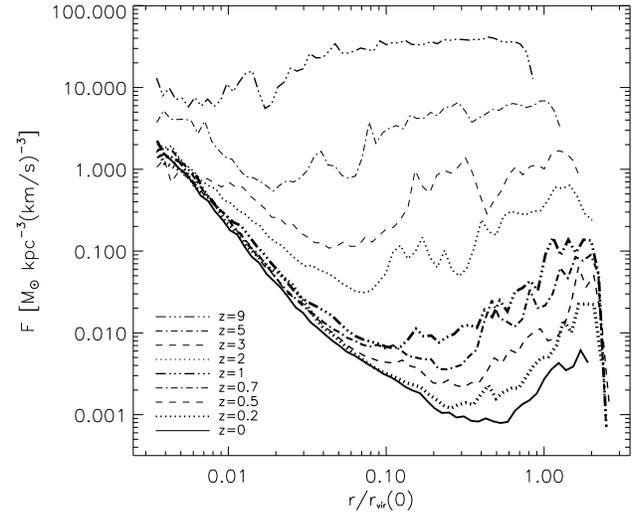,width=8.5cm}}
\caption{$F$ as a function of co-moving radius $r$ for all material
that lies within 2$r_{\rm vir}$ of the center of the G1
halo at $z=0$. Each curve corresponds to a different
redshift as indicated in the legends.}
\label{fig:f_allz}
\end{figure}

Figure~\ref{fig:f_allz} plots the median value ($F$) of the phase
space density $f$ as a function of distance to center of the most massive
progenitor at several different redshifts. At the highest redshift
plotted ($z=9$) the overall value of $F$ is higher at large radii than
it is at the center of the halo and only varies by a factor of few
over the entire range of radii ($F \sim 10-30\funits$).  This is
because within the inner, virialized regions the coarse-grained phase
space density is lowered due to mixing, while it is still high in the
outer regions where the large fraction of matter is not yet mixed or is located
in small subhalos, which have not mixed to the degree of the main host, and
the tidal streams they produce when disrupted.  As
the evolution progresses, there is a decrease in the central value of
$F$ till $z= 3$.  The median $F$ at the center of the halo drops from
$F \approx 10\funits$ at $z=9$ down to $ F \approx 2\funits$ at
$z=0$. Beyond $z=3$, the central value remains constant while there is
a steady decrease in $F$ at larger radii leading to the development of
an approximately power-law profile at $z=0$.

Although it is instructive to plot the mean and/or median values of
$F$ at each radius, much more information is contained in the full
coarse-grained phase-space density $f$.  The six-dimensional function
$f(\mathbf{x}, \mathbf{v})$ can be most easily visualized using the
volume DF $V(f)$ which also obeys a Mixing Theorem \citep{mathur_88}.
\citet{Arad04} showed that at $z=0$, $V(f)$ for cosmological dark
matter halos follows a power-law profile $V(f) \propto f^{-\alpha}$
with power-law index $\alpha = 2.4$ over 4 orders of magnitude in
$f$. They argued that since DM halos are almost spherical, if their
phase-space DFs are approximately isotropic their DFs could be written
as functions of energy alone: $f = f(E)$. In this case they showed
that if $Q \propto r^{-\beta}$, then $V(f)$ would also be described by
a power-law, and the index $\alpha$ was related to the index $\beta$
through a simple equation.

\begin{figure*}
\centerline{\psfig{figure=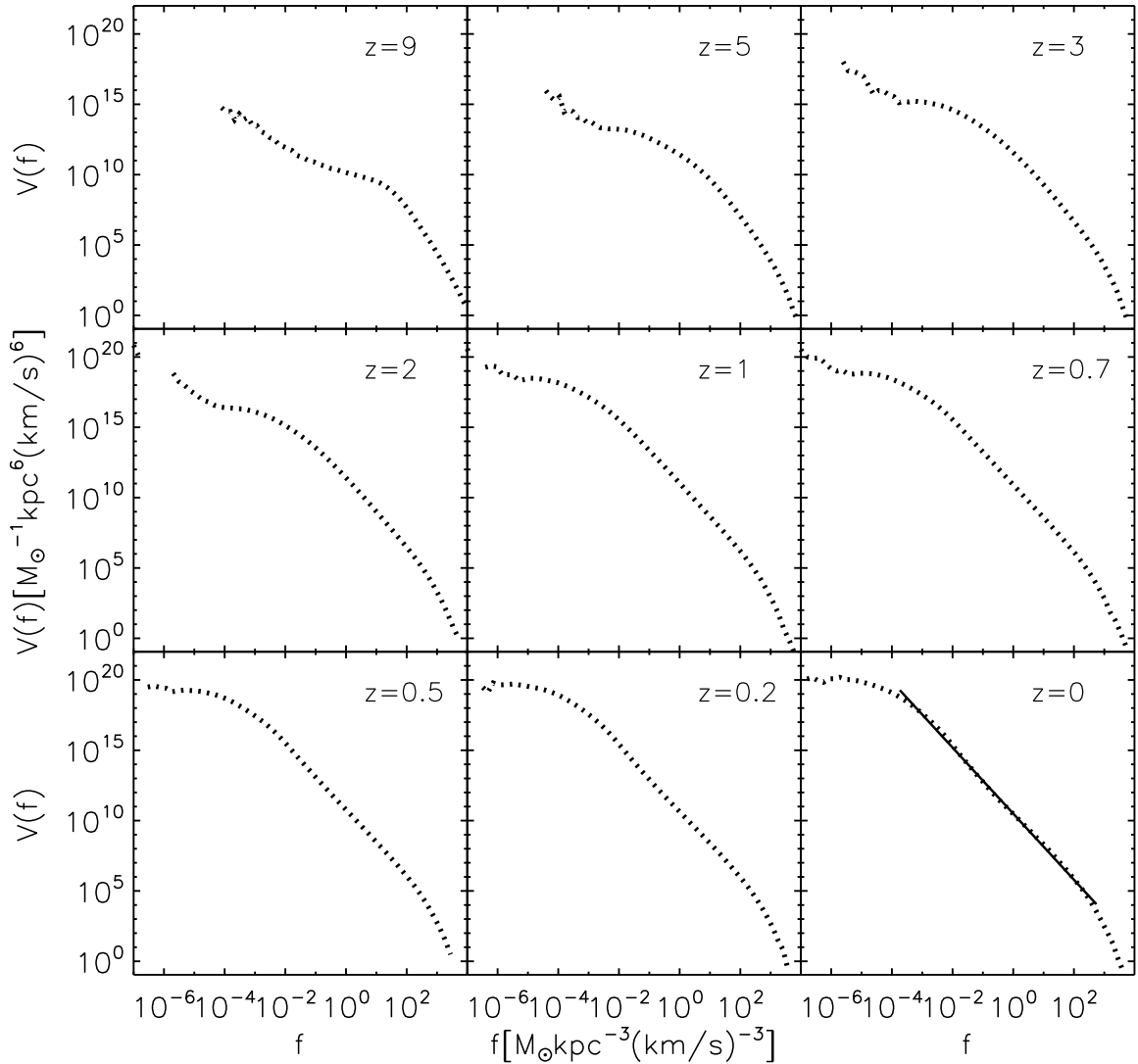,width=16.cm}}
\caption{The volume DF $V(f)$, for different redshifts in  the evolution of the
G1 halo in the L25 run.  A power-law fit $V(f) \propto f^{-2.34 \pm 0.017}$,
the best fit over this range of $f$, is shown at $z=0$ (solid line).
\label{fig:vf_L25}}
\end{figure*}

In Figure~\ref{fig:vf_L25} we plot the evolution of $V(f)$ with
redshift for all the material that lies within twice the
virial radius of halo G1 at $z=0$. 
In the bottom right hand panel, the solid line is a fit to $V(f)$ for values of
$10^{-4} <f < 10^{2.5}$. Over this range the best fit power-law profile is
given by $V(f) \propto f^{-2.34 \pm 0.02}$ which is close to the power-law
profile fit obtained by \citet{Arad04}.  At all redshifts $V(f)$ deviates from
the power-law profile at both low and high end. It is likely that at the
high-$f$ end the distribution deviates from the power-law due to the unresolved
subhalos below the mass resolution limit of the simulation. 

The fine-grained version of the volume $\mathbf{V}(\mathbf{f})$ of
phase-space associated with material of phase-space density
$\mathbf{f}$ is conserved by the collisionless Boltzmann equation,
however the coarse-grained version of this quantity ($V(f)$) always
{\it increases} by the {\it Mixing Theorem} \citep{mathur_88}.  We see
that increase in $V(f)$ occurs quite steadily from $z=9$ onward at all
values of $f$, but particularly for lower values of $f$. As the system
evolves there is more volume associated with material with low $f$. At
any value of $f$ the volume $V(f)$ increases with decreasing redshift.
Thus the behavior of the evolution of $f$ with redshift is completely
consistent with our expectations from the Mixing Theorems. While the
Mixing Theorems have previously been demonstrated for isolated
systems, this is to our knowledge the first demonstration of their
validity in a cosmological context.

\begin{figure*}
\centerline{\psfig{figure=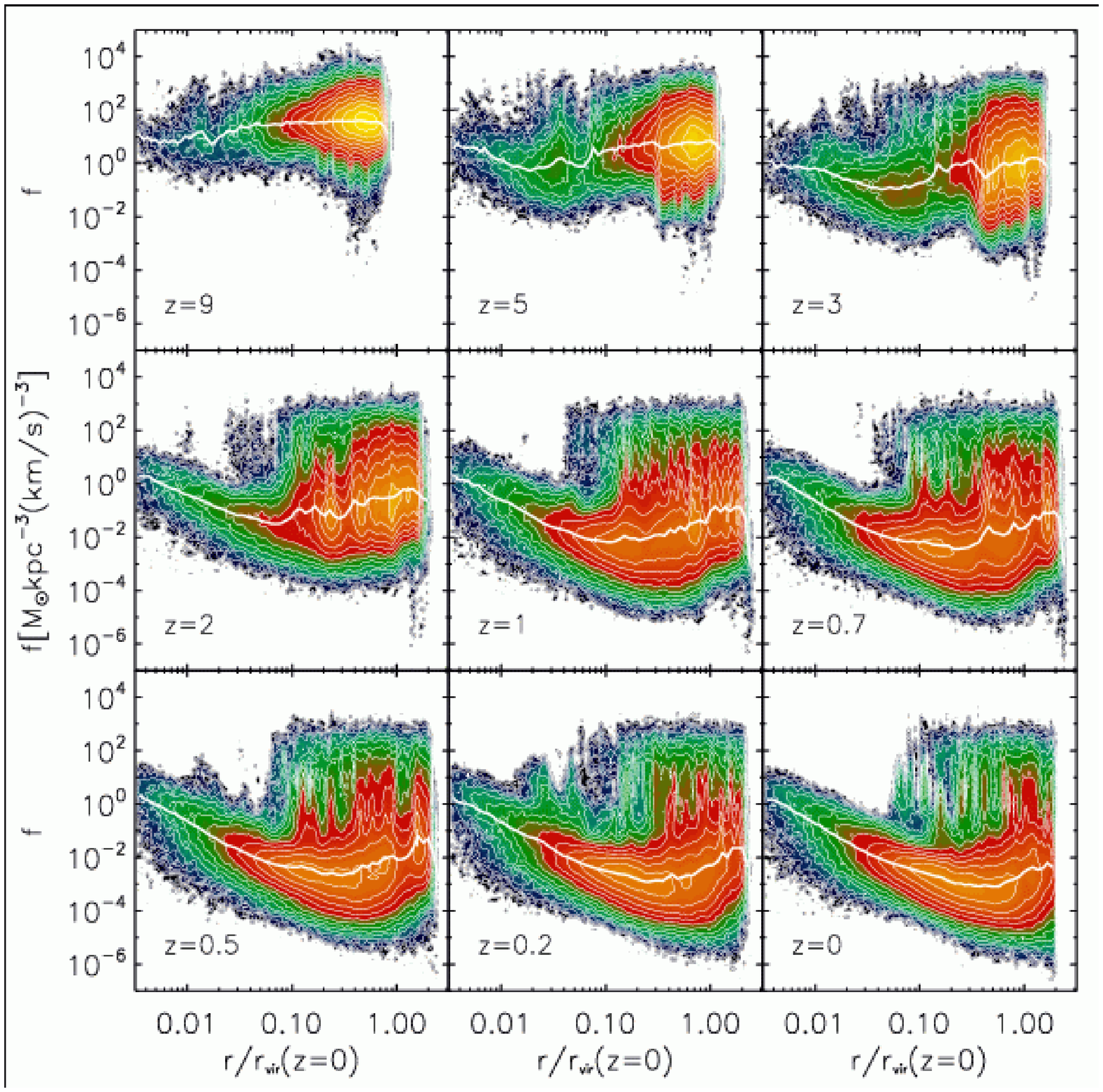,width=16.cm}}
\caption{Contours of constant particle number density in the plane
($\log(f), \log(r)$) for the G1 halo, plotted at different redshifts.
At each redshift the radial distribution of particles is given
relative to the most bound halo particle in units of the virial radius
of the halo $z=0$. The yellow and black contours correspond to the
regions with the highest and lowest particle number densities
respectively.  Contours are spaced at logarithmic intervals in
particle number density relative to the maximum density contour.  $F$,
the median value of $f(\mathbf{x}, \mathbf{v})$ is represented by the
solid white line.
\label{fig:rf_L25_cont}}
\end{figure*}

It is particularly illuminating to plot the full coarse-grained phase
space density of all particles as a function of radius from the center
of mass of the main halo at each redshift
(Figure~\ref{fig:rf_L25_cont}).  At any radius from the center there
exist particles with phase-space densities spanning between 4-8 decades in
$f$. The colored contours represent a constant particle number per
unit area on the plane $(\log(f), \log(r))$.  The yellow/orange
contours represent the parameter range with the largest number of
particles while the blue/black contours represent regions with the
smallest number of particles.  The solid white curves on each plot
show the median value ($F$) of $f(\mathbf{x}, \mathbf{v})$  in 100
logarithmically spaced bins in $r$. These curves are identical to the
various curves in Figure~\ref{fig:f_allz} and trace the regions with
the largest particle number density at each radius.  Numerous spikes
in $f$ are seen at $r > 0.1r_{\rm vir}$ and correspond to DM
sub-halos.

As the evolution proceeds, there is an overall lowering of the median
phase-space density $F$ (white curves).  There is also a continuous
decrease of the low-end envelope of $f$ to lower values pointing to an
increasingly large amount of matter with low phase-space
density. However the upper envelope representing the highest phase
space density regions at the centers of the subhalos remains
relatively unchanged with redshift indicating that primordial values
of $f$ are largely preserved at the centers of DM subhalos that lie
outside the center of the main halo.

\begin{figure*}
\centerline{\psfig{figure=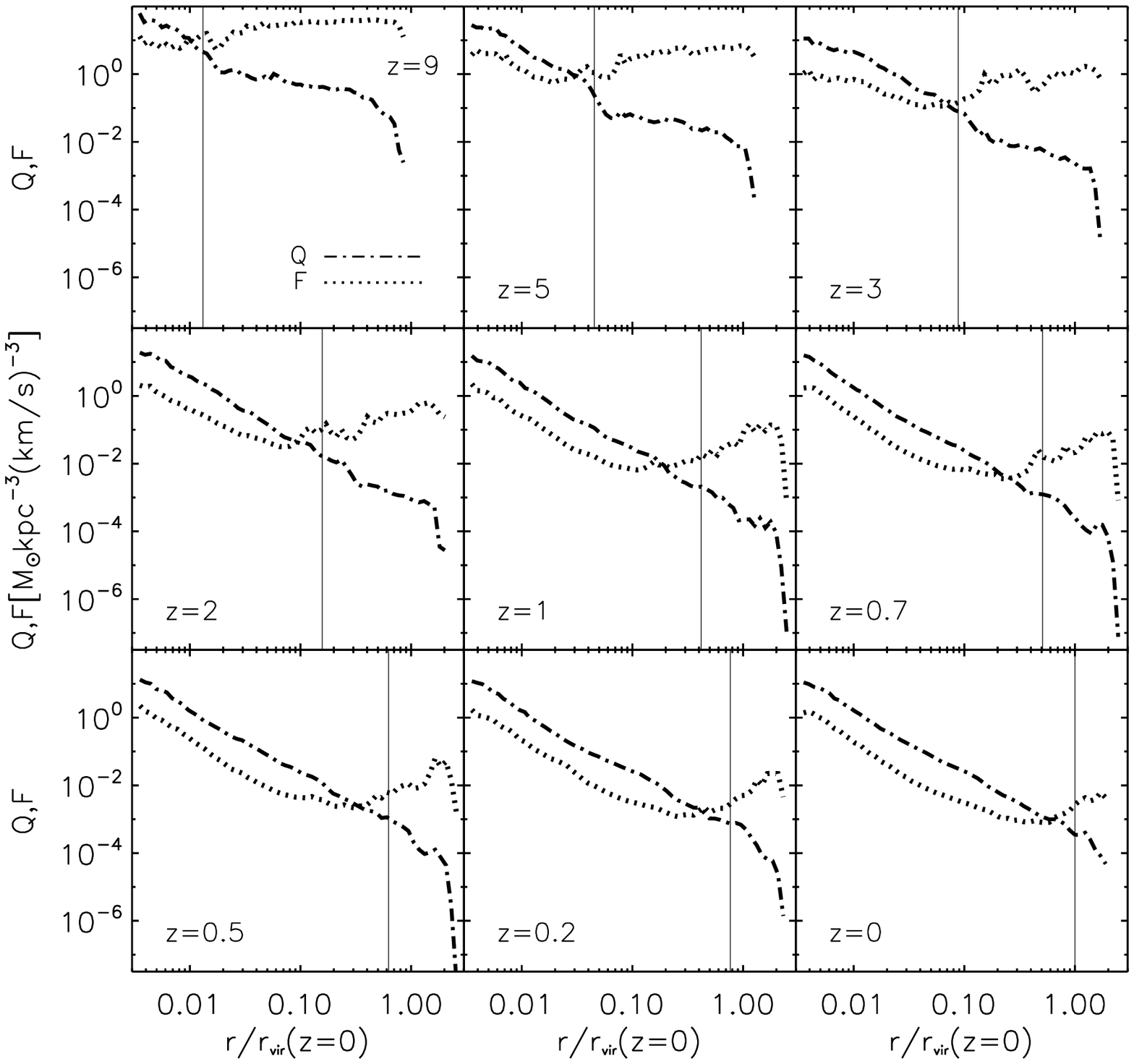,width=16.cm}}
\caption{$Q=\rho/\sigma^{3}$ (dot-dashed line) and median $F$ (dotted line) as
a function of $r/r_{\rm {vir}} (z=0)$ for different redshifts for the G1 halo.
The dashed vertical line is situated at the virial radius of the virialized
main halo at each redshift.
\label{fig:qf_L25}}
\end{figure*}

It is illustrative to compare the evolution of $F(r)$ and $Q(r)$ with
redshift.  Figure~\ref{fig:qf_L25} shows the two curves plotted as a
function radius (in co-moving units) at each redshift for all the
matter in halo G1 that lies within two virial radii at $z=0$, as a
function of physical radius from the center of the most massive
progenitor of the final halo (in units of the virial radius at $z=0$).
In each panel a thin vertical line is drawn at {\it the virial radius
of the halo at that redshift}.  We note that as expected from the
expansion of the Universe and increase in clustering, the virial
radius of the halo increases with decreasing redshift.

We find the best fit power-law $Q \propto r^{-1.84 \pm 0.012}$ at $z=0$ to the
profile within $0.6r_{\rm vir}$. In agreement with \citet{hoffman07}, we find
that the same power-law provides a reasonably good fit to the profiles at
redshifts from $z\sim 5$ to $z=0$.  While $Q$ always decreases with radius
since it is a spatial average over increasingly large volumes of configuration
space, the median phase-space density $F$ decreases monotonically with radius
only within the virial radius of the main progenitor at that redshift. Outside
the virial radius at each redshift (i.e. to the right of the vertical dashed
line), $F(r)$ become significantly flatter, even increasing with increasing
radius. At all $z \ne 0$ this represents the median phase-space density of
matter that will lie within $\sim 1$ virial radius at $z=0$, but is either
still unvirialized or lies within small halos which have not mixed to the
degree of the more massive main progenitor.  The nearly constant value of $F$
beyond the virial radius at high $z$ may be interpreted as the median phase
space density of DM in the Universe at that redshift. At lower redshifts, a
non-negligible fraction of this matter lies within inner regions of small
halos, which have undergone relatively small amount of mixing. As more and more
material undergoes substantial mixing as the halos evolve, the high phase-space
density unmixed material in the subhalo centers, as well as in the relatively
unmixed  streams formed from disrupted subhalos, prevents the median at large
radii from decreasing rapidly, despite the large increase in low material with
low $f$-values within the virialized regions of the main halo.

\begin{figure*}
\centerline{\psfig{figure=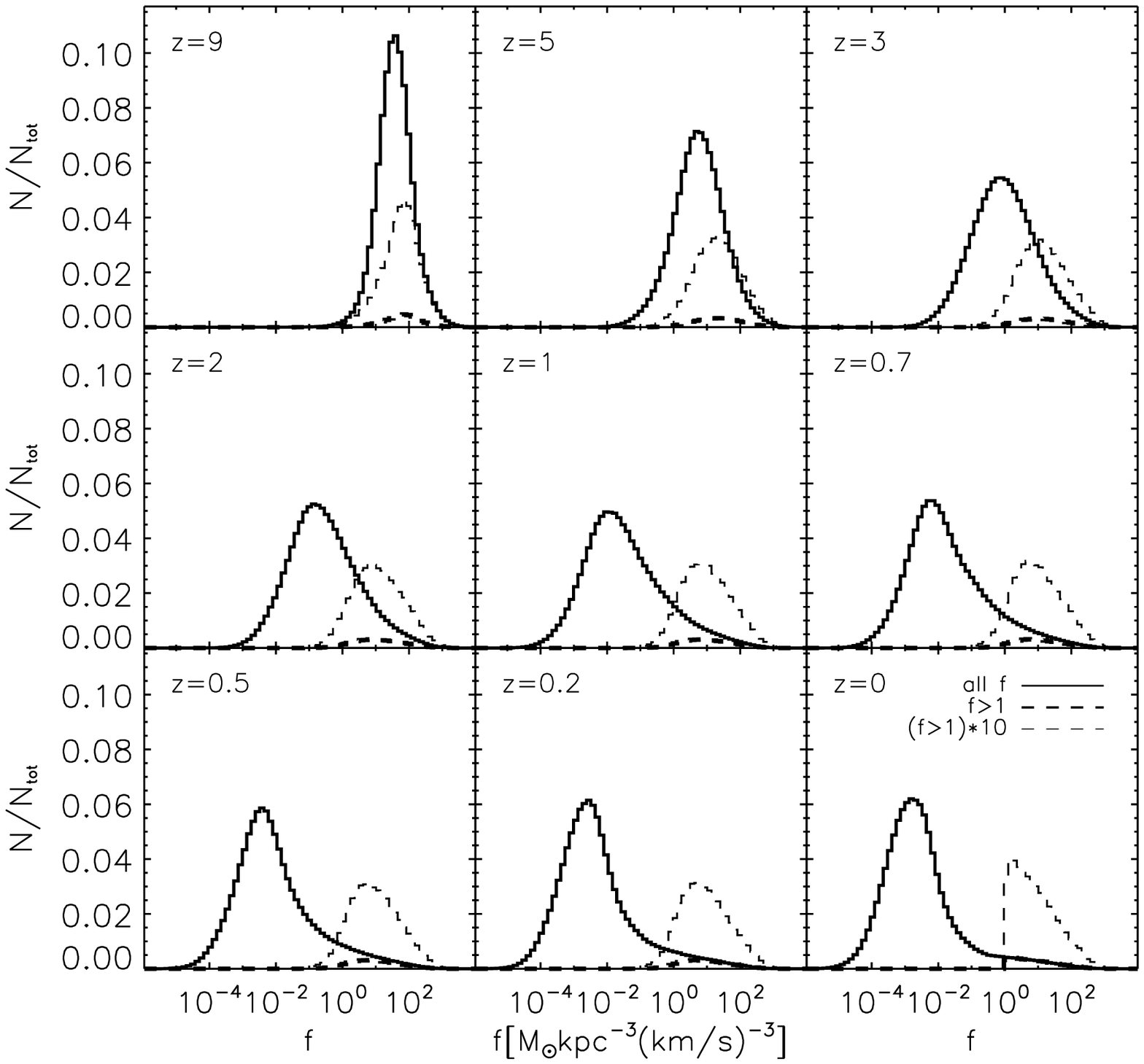,width=16.cm}}
\caption{Histograms of the phase-space density $f$ of all halo
particles as a function of redshift (solid black curves).  The thick
blue dashed curves follow those particles which have the highest
phase-space densities $f \ge 1$ at $z=0$. (The thin blue dashed line
is the same as the thick-dashed line, except that the corresponding
$y$-axis values multiplied by 10 to enhance visibility.)
\label{fig:hist_L25}}
\end{figure*}

To better understand the evolution of $f$ with redshift we plot
histograms of $\log(f)$ at each redshift in Figure~\ref{fig:hist_L25}
(solid line).  The initial distribution of $\log(f)$ at $z=9$ is close
to Gaussian (indicating that $f$ itself has a log-normal
distribution).  The peak of the Gaussian (the median of $\log(f)$)
moves to lower values as the halo evolves from $z=9$ to $z=0$ due to
both mixing in the main halo and in subhalos and due to cosmic
expansion.

By approximately $z = 1$, the skewness of the distribution is
significant because more and more matter undergoes mixing during halo
mergers and moves to lower values of $f$, and the skewness grows
steadily until $z=0$. The figure shows, however, that some fraction of
the high-$f$ matter survives to $z=0$.  The thick dashed line shows
the distribution of dark matter particles that have the values of $f
\ge 1\funits~$ (situated in the high $f$ tail of the histogram) at $z=0$. The
thin blue dashed line plots the same distribution but with $y$-axis
values multiplied by 10 to make the distribution more visible. As we
saw in Figure~\ref{fig:rf_L25_cont}, most of these high-$f$ tail
particles are in subhalos at $z=0$, while some are also in the highest
phase-space density particles in the central cusp of the halo at
$z=0$. This high-$f$ sub-population has a Gaussian distribution at
$z=9$ with a mean $f=1.73\pm0.6\funits~$  compared with the
mean $f=1.57\pm0.54\funits$  for all the particles (in the
solid curve) at $z=9$.  A Students' T-test indicated with 99\%
confidence that both distributions are drawn from the same population
at $z=9$. This implies that the material that lies in the centers of
DM subhalos at $z=0$ will have phase-space densities that are
representative of the mean phase-space density of matter at $z=9$. We
expect that at even higher redshifts, both distributions in $\log(f)$
will become much more sharply peaked asymptoting to a narrow function
at the start of the simulation, whose width represents the true range
of $f$ for CDM particles.

This decrease in $F$ of ``unprocessed'' material in Figure~\ref{fig:qf_L25} and
the decrease (see Fig.~\ref{fig:hist_L25}) of peak of $\log(f)$ is a result of
two independent factors. First, mixing (both due to merging between subhalos
and due to merging of subhalos with the main halo), causes a decrease in the
overall coarse-grained phase-space density, as anticipated from the Mixing
Theorem.  Second, at most redshifts, a significant fraction of matter has not
yet been virialized or is witin much smaller halos and lies outside the virial
radius at that redshift (to the right of the vertical line in
Fig.~~\ref{fig:qf_L25}). The density of this unprocessed material decreases due
to the cosmic expansion as ($\bar\rho(z) \propto
(1+z)^3$)\footnote{Hubble expansion does not contribute to a change in the
local velocity dispersion of dark matter particles that contributes to phase
space density.}. Thus the decrease in median phase-space density of a halo with
redshift is the consequence of both relaxation and mixing within virialized
halos and subhalos, and the cosmic expansion of the Universe.

\bigskip
\section{Comparsion of four Milky-Way sized halos}
\label{sec:otherhalos}

In this section we compare the evolution of phase-space DFs for all
four of the DM halos described in \S~\ref{sec:numerical}.

\begin{figure}
\centerline{\psfig{figure=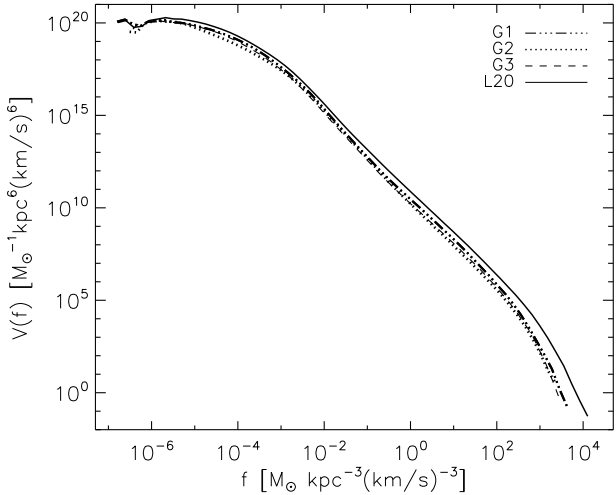,width=8.5cm}}
\caption{$V(f)$ for four halos, at $z=0$.  The triple-dot-dashed
line is for $G1$, the dotted line for $G2$, the dashed line for $G3$, and the
solid line is for higher mass resolution simulation $L20$.
\label{fig:vf_halos}}
\end{figure}

Figure~\ref{fig:vf_halos} compares the volume distribution of phase
space density $V(f)$ for the four different halos at $z=0$.  All three
halos from the L25 simulation (G1, G2, G3) show almost identical
profiles in $V(f)$ confirming the universality of the process that
produced the phase-space DF. For halo L20, $V(f)$ lies systematically
above the other curves especially at higher values of $f$, where it
also extends to large values of $f$. This is numerical consequence of
the increased mass resolution of the simulation as was first shown by
\citet{shar_stein06}. This indicates that the absolute value of $f$
derived in the previous section is somewhat dependent on the mass
resolution of the simulation and increases slightly with increasing
mass resolution.  In all four halos $V(f)$ is well approximated by a
power-law over nearly 6 orders of magnitude in $f$. The first column
of Table~\ref{tab:halos_fits} gives details of the power-law fits to
$V(f)$ (i.e. the power-law slopes and their errors over the range
$10^{-4} < f < 10^{2.5}$.) The power-law slopes we obtained are
similar to those obtained by previous authors \citep{Arad04}.

\begin{figure}
\centerline{\psfig{figure=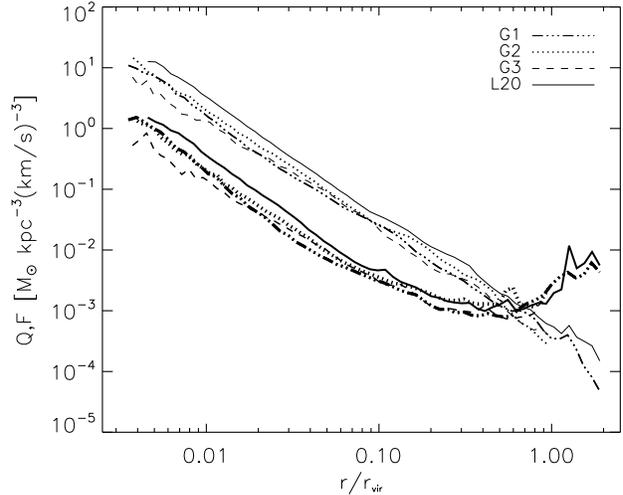,width=8.5cm}}
\caption{$Q$ (thin lines) and $F$ (thick lines) for all the halos studied, at
$z=0$.  Triple-dot-dashed lines are for $G1$, dotted lines for $G2$, dashed
lines for $G3$, and the solid lines are for $L20$.
\label{fig:qf_halos}}
\end{figure}

In Figure~\ref{fig:qf_halos} we plot $F(r)$ and $Q(r)$ for the four
different halos at $z=0$. The top set of four curves show $Q(r)$ while
the lower set of curves show $F(r)$.  In all four halos $Q(r)$, is
well fitted by a power-law of slope $\beta = 1.8 - 1.9$ (see
Table~\ref{tab:halos_fits}).  $F(r)$ shows significant deviations from
a simple power-law profile, with a systematic upturn beyond $r/r_{\rm
vir} > 0.1$.  The higher resolution simulation L20 appears to have
systematically higher $F$ and $Q$ values than the halos from the low
resolution simulation (a numerical consequence of the higher mass
resolution, \citet{shar_stein06}.) Table~\ref{tab:halos_fits} gives
the values for the slopes and the error-bars on the power-law fits to
$Q(r)$ and $F(r)$ for $r < 0.6 r_{\rm vir}$ as well as the inner
power-law slope of $F(r)$.

\begin{table*}
\centering
\caption{Power-law indices for fits to $V$, $Q$, and $F$ for four halos at
$z=0$ for $r < 0.6r_{\rm vir}$ }\label{tab:halos_fits}
\begin{tabular}{@{}lccc}
\hline
Halo & $V$ & $Q$ & $F$ \\
\hline
G$_{1}$ &$-2.34 \pm 0.02$  & $-1.84 \pm 0.01$ & $-1.59 \pm 0.05$  \\
G$_{2}$ &$-2.37 \pm 0.02$  & $-1.82 \pm 0.01$ & $-1.46 \pm 0.06$  \\
G$_{3}$ &$-2.35 \pm 0.02$  & $-1.75 \pm 0.01$ & $-1.42 \pm 0.03$  \\
L20         &$-2.27 \pm 0.01$  & $-1.87 \pm 0.01$ & $-1.64 \pm 0.07$  \\
\hline
\end{tabular}
\end{table*}

\bigskip
\section{On the interpretation  of  $Q(r)$ and phase-space density distributions}
\label{sec:origin}

Several previous studies have attempted to account for the origin of
the power-law $Q(r)$ profiles of DM halos.  Notably, it has been
argued that power-law profiles result from virialization and not from
the hierarchical sequence of mergers, since they are also produced in
simple spherical gravitational collapse simulations
\citep{taylorN01,barnes06, barnes07}. Our results presented in the
previous section (Figure~\ref{fig:qf_L25}) show that while $Q(r)$ and
$F(r)$ have approximately power-law form within $0.6 r_{\rm vir}$ at a
given redshift, $F(r)$ flattens out and remains quite flat beyond this
radius. Furthermore, as the hierarchical growth of the halo progresses
these approximately power-law profiles extend to larger radii until
they encompass all the mass within the virial radius at $z=0$.  In
this section we attempt to understand the origin of approximately
power-law profiles.

It has been shown from both theoretical arguments and numerical
simulations \citep{dehnen2005,vass_etal_08a} that power-law profiles
of phase-space density for central cusps are well preserved during
major mergers. At intermediate times during a major merger, deviations
from the power-law profile are seen but these largely disappear at the
end of the merger. In major mergers there are two main reasons for the
preservation of the power-law profiles. First, the most tightly bound
material - that forming a steep central cusp or shallow core preserves
its phase-space density in the final remnant. This is a consequence of
the additivity of the excess mass function - a result of the fact that
steeper cusps are less mixed than shallower cusps \citep{dehnen2005}.
Over 60\% of the material in the central cusp (within one scale radius) of a
progenitor NFW halo remains within the cusp of the merger remnant
\citep{valluri07}.  The second reason for preservation of power-law
profiles at large radii is that material outside 3 scale radii of the
progenitor halos is redistributed to other radial bins almost
uniformly from each radial interval. In addition nearly 40\% of the
material within the virial radius of the progenitor halos is ejected
to beyond the virial radius of the final remnant
\citep{kazantzidis_etal06, valluri07}, and this material can be shown
to have originated in roughly equal fractions from each radial
interval beyond 3 scale radii and has higher phase-space density than
expected from the simple power-law extrapolation of the inner
power-law. This self-similar redistribution of material
contributes to the preservation of power-law profiles in
$Q(r)$ and $F(r)$ in equal mass binary mergers. Thus, once a power-law
phase-space DF has been established in a DM halo major mergers will
not destroy this profile.

As was discussed in the previous section (Figure~\ref{fig:hist_L25}),
the coarse-grained phase-space density $f$ in $\Lambda$CDM halos has a
log-normal distribution with the median of $\log(f)$ (the peak of the
histogram $f_{\rm peak}$) evolving steadily toward lower values of $f$
as the halo grows.  This evolution results from both the decrease in
$f$ due to mixing when matter accretes onto a halo and virializes and
due to the decrease in the mean density as the Universe expands
(cosmic expansion reduces the mean density of the Universe as
$\bar{\rho}(z) \propto (1+z)^{3}$.) Decrease in the median phase-space
density $f_{\rm peak}$ with decreasing redshift should therefore be at least as fast as the decrease in the mean density of the Universe, or faster. The decrease of phase-space
density $f$ due to mixing is the consequence of complex non-linear
evolution, hence it is difficult to arrive at a analytic expression to
represent the cosmological evolution of the median phase-space density $f_{\rm
peak}$. We therefore attempt to derive it empirically.
 
\begin{figure}
\centerline{\psfig{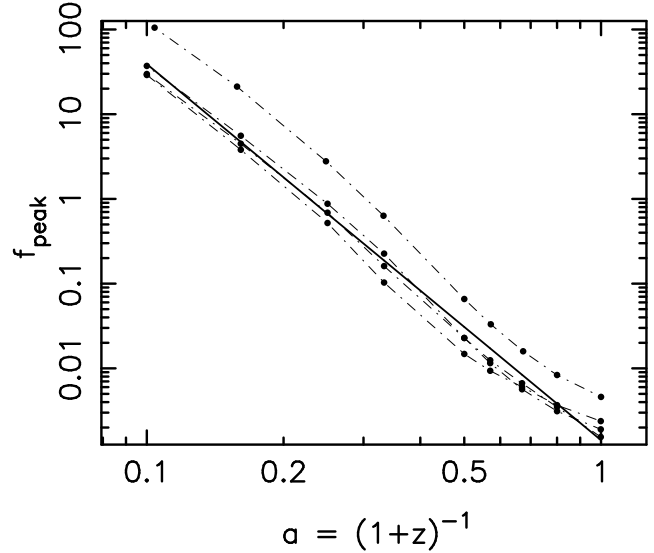}}
\caption{The median phase-space density (peak of histograms in
Fig.~\ref{fig:hist_L25}) ($f_{\rm peak}$) at each redshift as a
function of the expansion parameter $a= (1+z)^{-1}$ for four Milky-way
sized halos (points). The thin dot-dashed lines connect points for
individual halos.  The best-fit (solid line) has a slope of $-4.3\pm
1.1$.
\label{fig:medf}}
\end{figure}

Figure~\ref{fig:medf} shows the median phase-space density
$f_{\rm peak}$ at each redshift derived from the histograms of
$\log(f)$ in Figure~\ref{fig:hist_L25} for each of the 4 Milky-Way
sized cosmological halos in this study.  The points represent the
values of $f_{\rm peak}$ as a function of expansion factor $a=(1+z)^{-1}$, while
the thin dot-dash lines connect points for each of the 4 halos. The
thick solid line is the best fit power-law for all halos taken
together: $f_{\rm peak}(a) \propto a^{-4.3\pm 1.1}$. As expected, this
decrease is faster than $a^{-3}$.

Thus on average, as a halo grows via accretion its median density
decreases as a power-law with time, despite the fact that the most
centrally concentrated material retains its original high phase-space
density. This power-law profile in $f_{\rm peak}$ leads to some
insights on the development of phase-space density profiles.  We can
break down the formation of halos into two phases \citep{li_etal07}:
the fast accretion regime during which halo mass grows very rapidly
and the slow accretion regime.  In the fast accretion regime, the
mixing processes are very efficient as the potential well is
established and potential fluctuates rapidly and constantly. This
results in a rapid decrease in the over all central phase-space
density of the halo (as seen from the rapid drop in the central most
regions of the profiles in Fig.~\ref{fig:f_allz}.)  The inner profile
of phase-space distribution should be largely set during this stage.
As halos grow subsequently, either by major mergers or quiescent
accretion of smaller halos, the average phase-space density decreases,
but the pre-existing high central phase-space density cusps are are
preserved.

At these later times the evolution of the true phase-space density is
complex and occurs due to the accretion of high phase-space density
material in subhalos as well losely bound material at the edges of the
halo. The steady decrease in the amplitude and increase in the slope of the $F(r)$
profiles in Fig.~\ref{fig:f_allz} with redshift show that the true phase-space
density does not obey the  power-law profiles seen in $Q(r)$
at all redshifts.  Variations in the individual power-law profiles of
halos both at $z=0$ (Fig. ~\ref{fig:qf_halos}) and with redshift
reflect the large variation in the cosmic accretion histories of
individual halos. Additional deviation is due to the matter bound to
subhalos that survives with high phase-space density and leads to the
variation of the coarse-grained $f$ of some six orders of magnitude in
the outer regions of halos.

All this indicates that the actual phase-space distribution is not as
universal and simple as the $Q(r)$ profiles lead one to believe.  The
mechanism behind the universal power-law form of the $Q(r)$ profiles
is therefore likely to be different (e.g., it is not affected by the
presence of subhalos) and simpler than the processes that shape the
distribution of the coarse-grained phase-space density.  The $Q$
profile is a ratio of the density and velocity dispersion (which can
be interpreted as a measure of temperature) and is therefore related
to the entropy. For an ideal monatomic gas the entropy can be defined
as $K_{\rm gas}=T/\rho^{2/3}$, while for dark matter, by analogy, the
entropy can be defined as $K_{\rm dm}=\sigma^2/\rho_{\rm dm}^{2/3}$
\citep{faltenbacher_etal07}.  This, as noted by \citet{hoffman07},
gives $K_{\rm dm}\propto Q^{-2/3}$ or $K_{\rm dm}\propto r^{1.2}$ for
$Q\propto r^{-1.8}$.  This power-law form and slope of the entropy
profile is very similar\footnote{The slope is even more similar if one
takes into account the random bulk motions of the gas in estimating
$K_{\rm gas}$ \citep{faltenbacher_etal07}.} to the one found for the
gas in the outer regions of clusters in cosmological simulations
\citep[e.g.,][]{borgani_etal04,voit_etal05}.  This power-law is also
predicted by the models of spherical accretion \citep{tozzi_norman01}
and reflects the encreasing entropy to which the accreting material is
heated as the halo grows its mass. Although the processes governing
the virialization of gas and dark matter are different (the
short-range local ineractions for the former, and long-scale
interactions for the latter), the fact that the resulting entropy
profiles are quite similar indicate that they lead to the same
distribution of entropy. The $Q(r)$ profile therefore may reflects the
overall entropy profile of dark matter, not the coarse-grained local
phase-space density, which exhibits a more complicated behavior.

\bigskip
\section{Summary and Conclusions}
\label{sec:conclude}

We have investigated the evolution of the phase-space density of the
dark matter in cosmological simulations of the formation of Milky-Way-
sized dark matter halos. The analysis was carried out using two
different codes for estimating the phase-space density. Both codes
give qualitatively similar results, but the estimated values of phase-
space density $f$ are quite sensitive to the type of code and, for a
given code, also depend quite sensitively on the choice of smoothing
kernel used. Based on comparisons of the two codes (\fiestas and
\enbid) and various smoothing parameters (Appendix), we select the
\enbid code with $n=10$ kernel smoothing and present results for the
analysis with this set of parameters.

The simulations presented in this paper complement our analysis of the
evolution of phase-space density in binary major mergers
\citep{vass_etal_08a}.  We confirm that the profiles of
$Q(r)=\rho_{\rm dm}/\sigma^3_{\rm dm}$ computed by previous authors
can be described by a power-law $Q(r)\propto r^{-1.8\pm 0.1}$ over
more than two orders of magnitude in radius in all halos.  The median
of the phase-space density ($F(r)$) at given radius $r$, however,
exhibits a more complicated behavior. Although $F(r)$ is approximately
a power-law for $r< 0.6r_{\rm vir}$, the profiles generally flatten in the outer
regions. Subhalos contribute somewhat to this behavior, although their
effect is limited by the relatively small fraction of mass ($<0.1$) bound to
them. However, in addition to subhalos, a significant fraction of high 
phase-space density matter is in the relatively unmixed streams of 
already disrupted subhalos \citep[e.g.,][]{Arad04,diemand_etal08}. The fraction of mass
in such streams can be substantial at large radii. 

This behavior holds at earlier epochs.  From $z=5$ to $z=0$, material
within $r_{\rm {vir}}(z)$ at each redshift follows a power-law in $Q$
with an approximate power-law slope of $\sim -1.8$ to $-1.9$. In
contrast, $F(r)$ can only be well described by a power-law in the inner
regions, and its slope changes continuously with redshift. Beyond the
virial radius, $Q$ (a quantity that is obtained by averaging over
increasingly large volumes) decreases rapidly with radius, but the
median value of $F$ flattens significantly.  We argue that $F$ is a
more physically meaningful quantity, especially for understanding the
evolution of phase-space density in collisionless DM halos, as it
measures the median of the true coarse-grained phase-space density.

At all redshifts, the highest values of phase-space density $f$ are
found at the centers of dark matter subhalos.  In the center of the
main halo, the median phase-space density ($F$) drops by about an
order of magnitude from $z=9$ to $z=0$.  In contrast, the centers of
DM subhalos maintain their high values of $f \sim 10^3\funits$ at all
redshifts. The highest values of $f$ at the center of the main halo are,
therefore, lower and less representative of the primordial phase-space
density of dark matter particles than the central value of $f$ in the
high phase-space density subhalos. At $r_{\rm {vir}}$, the decrease in
median phase-space density is much more significant, with $F \approx
30\funits$ at $z=9$ decreasing to $ F\approx 10^{-3}\funits$ at
$z=0$, a decrease of over four orders of magnitude
(Figure~\ref{fig:rf_L25_cont}).

The evolution of $F(r)$ and $V(f)$ with redshift are consistent
with expectations from the Mixing Theorems, which require that mixing
reduces the overall phase-space density of matter in collisionless
systems and that the volume of phase-space associated with any value
of $f$ increases due to mixing and relaxation.
 
The distribution of $f$ is approximately log-normal until $z\sim
3$. As time progresses, the mean and median of $\log(f)$ shift to
progressively lower values as a larger and larger fraction of matter
undergoes mixing and moves to lower values of $f$.  Some fraction of
high phase-space density material does survive in the centers of
subhalos and in the relatively unmixed streams leftover after subhalo
disruptions, which skews the distribution.  Remarkably, the highest
phase-space density particles at $z=0$ have retained their phase-space
density since $z \approx 9$, the earliest epoch we analyzed.  The
phase-space density in the centers of dark matter subhalos is
therefore representative of the mean phase-space density of DM at the
highest redshifts. This can potentially allow for stronger constraints
to be placed on the nature of DM particles from the Tremain-Gunn bound
\citep{tremaine_gunn_79, hogan_dalcanton_00}.

The median value of phase-space density $f_{\rm peak}$ decreases with
decreasing redshift and is given by the power-law $f_{\rm peak}(a)
\propto a^{-4.3\pm1.1}$ and is the result of two different processes:
(a) mixing within virialized halos which reduces the coarse-grained
phase-space density of matter that has turned around from the Hubble
flow and (b) the cosmic expansion of the Universe which results in a
decrease in the configuration-space density of ``unprocessed''
material (material that has yet to fall into a massive halo and become
virialized). Thus, as a halo grows by accretion and the Universe
expands, the median phase-space density in halos decreases steadily.

\bigskip

\section*{Acknowledgments}

We thank Y. Ascacibar and S. Sharma for the use of the \fiestas code and \enbid
code, respectively.  We especially thank S. Sharma for detailed discussions on
his \enbid code.  IMV acknowledges the support of National Science Foundation
(NSF) grant AST-0307351 to the University of Florida.  IMV, MV, AVK, and SK
were supported in part by Kavli Institute for Cosmological Physics at the
University of Chicago (where most of this work was carried out) through grants
NSF PHY-0114422 and NSF PHY-0551142 and an endowment from the Kavli Foundation
and its founder Fred Kavli.  AVK is also supported by the NSF under grants
AST-0239759 and AST-0507666 and by NASA through grant NAG5-13274. AVK would
like to thank the Kavli Institute for Theoretical Physics,  supported in part
by the NSF under grant PHY05-51164, for hospitality during the final editing of
the paper and participants of the workshop ``Back to the Galaxy'' for useful
discussions and feedback on the results of this paper.  SK is supported by the
Center for Cosmology and Astro-Particle Physics (CCAPP) at The Ohio State
University.

\bigskip

\appendix
\section{Comparison of  ``\fiestas''  and ``\enbid'' Analysis of Halo G1}

The numerical estimation of coarse-grained phase-space densities
during the evolution of the 4 N-body halos presented in this paper was
carried out using two publicly available codes ``\fiestas''
\citep{ascasibar_binney05} and ``\enbid'' \citep{shar_stein06}. A
comparison of these two codes has previously been presented by
\citet{shar_stein06} who showed that for an analytic DF (e.g. for the
spherical Hernquist potential), that ``\enbid with $n=10$ kernel
smoothing'' gave the highest fidelity to the analytical DF.  In a
related work \citet{vass_etal_08a} confirmed the findings of
\citet{shar_stein06} for a spherical isotropic NFW halo. This latter
study is the main basis for our choice of ``\enbid with $n=10$ kernel
smoothing'' .

However, it is unclear whether the comparisons with analytic profiles
of isolated halos at $z=0$ are valid for matter distributions arising
from cosmological N-body simulations, at high redshifts,
where the majority of the particles actually lie outside virialized
halos. Our purpose in this Appendix is to present a comparison of
results obtained with the different codes at a range of redshifts to
allow readers to appreciate how sensitive some of the results
presented in this paper are to the choice of code and smoothing
parameters used in density estimation.

\begin{figure}
\centerline{\psfig{figure=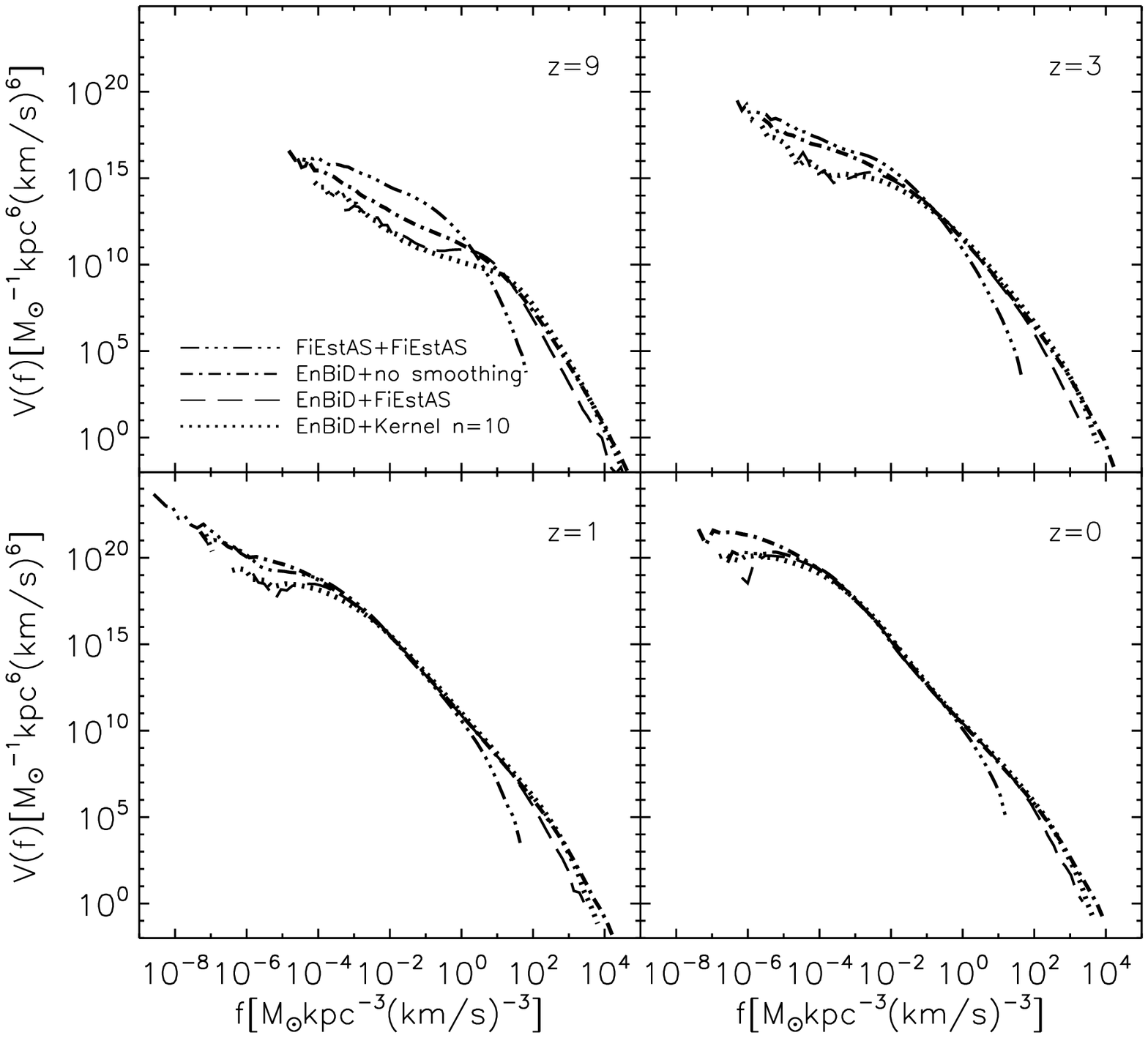,width=8.5cm}}
\caption{The volume DF $V(f)$, at different redshifts during the
evolution of the G1 halo in the L25 run, resulting from the \fiestas
and \enbid with three different parameters as indicated in the line
legends.
\label{fig:fvfz}}
\end{figure}

Since the coarse-grained DF $f(\mathbf{x}, \mathbf{v})$ is a
6-dimensional function, it is difficult to compare estimates for this
function obtained using different codes. Traditionally the best single
variable function to compare is the volume DF $V(f)$.  Variations in
the estimation of $f$ translate to variations in estimation of $V(f)$.
In Figure~\ref{fig:fvfz} we compare the volume density of phase-space
$V(f)$ obtained for the G1 halo at 4 different redshifts using the
\fiestas code (triple-dot-dashed curves), \enbid code with no smoothing
(dot-dashed curves), \enbid with \fiestas smoothing (long-dashed
curves), and \enbid with a $n=10$ kernel ( dotted curves). In
each case we see that $V(f)$ has a nearly power-law distribution over
more than six orders of magnitude in $f$ at $z=0$ (from
$\sim 10^{-4}$--$10^{3}\funits$). Table~\ref{tab:voffslopes} gives the slopes of power-law fits for the
four different estimates of $V(f)$ over the range $10^{-4} < f <
10^{3}$ at $z=0$. The \fiestas estimate of $V(f)$ is
systematically lower than all \enbid estimates at high values of  $f$ (at all redshifts).  In addition all \enbid curves extend to much higher values of $f$ than the \fiestas estimate (this is particularly true at $z=9$ where the \fiestas estimate differs from the other estimates both quantitatively and qualitatively).  The results from the various \enbid estimates differ very little at intermediate values of $f$ ($10^{-4}$--$10^{3}$) at $z=0$ and consequently we are confident that conclusions drawn from the median $f$ is quite insensitive to the details of the \enbid parameters used to obtain $f$.

\begin{table}
\centering
\caption{\label{tab:voffslopes} Power-law indices $V(f) \propto f^{-\alpha}$ and $F(r)\propto r^{-\beta}$ at $z=0$ from different codes}
\begin{tabular}{lcc}
\hline
Code                                      &  $\alpha$                & $\beta$       \\
\hline
\fiestas                                    & $2.62 \pm 0.06$  & $1.43\pm0.02$\\
\enbid  (no smoothing)        & $2.35 \pm 0.02$  & $1.65\pm0.02$\\
\enbid (\fiestas  smoothing)& $2.46 \pm 0.03$  & $1.69\pm0.07$\\
\enbid (kernel $n=10$)       & $2.34 \pm 0.02$  & $1.59\pm0.05$\\
\hline
\end{tabular}
\end{table}

\begin{figure}
\centerline{\psfig{figure=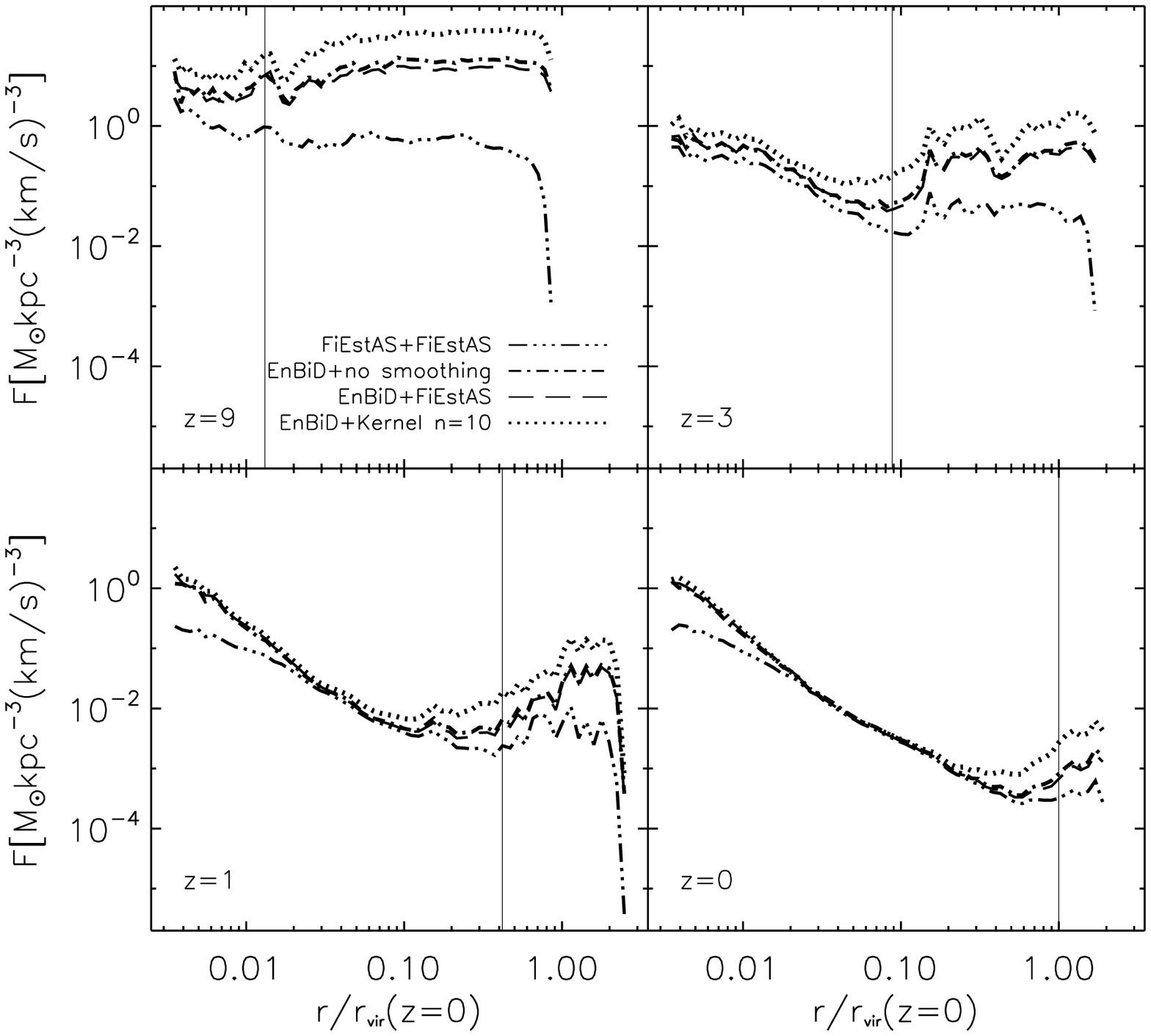,width=8.5cm}}
\caption{ $F$ for the G1 halo in the L25 simulation at different
redshifts obtained using four different analyses as indicated by
line-legends.  The dashed vertical line is situated at the virial
radius of the virialized main halo at each redshift.
\label{fig:allfzs}}
\end{figure}

The differences between the various estimates of $F$ at large radii become
signficantly larger at higher redshift. In Figure~\ref{fig:allfzs} we plot the
four different estimates of $F(r)$ at 4 different redshifts in the evolution. 
The vertical dashed line in each panel represents the virial radius of the main
halo at that redshift. We see that $F$ from any of the codes is quite flat
beyond the virial radius in all cases, but the absolute values of the curves
differ significantly. Note that at $z=9$, the different estimates can differ by
as much as two orders of magnitude. 

In this paper we choose to present results from \enbid with $n=10$
kernel smoothing over the other estimates largely because it does the
best job of reproducing analytic DFs \citep{shar_stein06} and because
it appears to provide a good upper limit to the phase-space density
both at high and low values of $f$. In the absence of an analytic comparison of the estimates at
high redshift, we caution the reader to refrain
from drawing very strong conclusions regarding absolute values of $f$
or $F$ from the results presented here.

\bibliographystyle{mn2e}
\bibliography{main2}

\bsp

\label{lastpage}

\end{document}

%% file: mycommands.tex
\def\HI{\ifmmode{\rm HI}\else{H\/{\sc i}}\fi}

\def\lsun{\ifmmode{{\mathrm L}_{\odot}}\else{L$_{\odot}$}\fi}

\def\msun{\ifmmode{{\mathrm M}_{\odot}}\else{M$_{\odot}$}\fi} 
\def\msunpc2{\ifmmode{{\mathrm M}_{\odot} \, {\mathrm{pc}}^{-2}}\else{M$_{\odot} \, {\mathrm {pc}}^{-2}$}\fi}

\def\kms{\ifmmode{{\mathrm{km \, s^{-1}}}}\else{${\mathrm{km \, s^{-1}}}$}\fi}